\documentclass[floats,floatfix,showpacs,amssymb,prd,superscriptaddress,twocolumn,aps]{revtex4-1}
\usepackage{graphicx, epsfig, amssymb} 
\usepackage{amsmath, amsfonts}
 \usepackage{gensymb}
\usepackage{bm} 

\usepackage[linktocpage]{hyperref}
\usepackage[caption=false]{subfig}
\usepackage[usenames]{color}

\def\be{\begin{equation}}
\def\ee{\end{equation}}
\def\beq{\begin{eqnarray}}
\def\eeq{\end{eqnarray}}

\newcommand{\bea}{\begin{eqnarray}}
\newcommand{\eea}{\end{eqnarray}}
\newcommand{\ben}{\begin{enumerate}}
\newcommand{\een}{\end{enumerate}}
\newcommand{\bi}{\begin{itemize}}
\newcommand{\ei}{\end{itemize}}

\pdfoutput=1

\begin{document}

\title{Effective stability against superradiance of Kerr black holes with synchronised hair}

 \author{Juan Carlos Degollado}
   \affiliation{
   Instituto de Ciencias F\'\i sicas, Universidad Nacional Aut\'onoma de M\'exico, Apdo. Postal
48-3, 62251, Cuernavaca, Morelos, M\'exico
 }

 \author{Carlos~A.~R.~Herdeiro}
   \affiliation{
   Departamento de F\'\i sica da Universidade de Aveiro and CIDMA, 
   Campus de Santiago, 3810-183 Aveiro, Portugal.
 }

 \author{Eugen Radu}
   \affiliation{
   Departamento de F\'\i sica da Universidade de Aveiro and CIDMA, 
   Campus de Santiago, 3810-183 Aveiro, Portugal.
 }


\date{February 2018}

\begin{abstract}
Kerr black holes with synchronised hair~\cite{Herdeiro:2014goa,Herdeiro:2016tmi} are a counter example to the no hair conjecture, in General Relativity minimally coupled to simple matter fields (with mass $\mu$)  obeying all energy conditions. Since these solutions have, like Kerr, an ergoregion it has been a lingering possibility that they are afflicted by the superradiant instability, the same process that leads to their dynamical formation from Kerr. A recent breakthrough~\cite{Ganchev:2017uuo} confirmed this instability and computed the corresponding timescales for a sample of solutions. We discuss how these results and other observations support two conclusions: $1)$ starting from the Kerr limit, the increase of hair for fixed coupling $\mu M$ (where $M$ is the BH mass) increases the timescale of the instability; $2)$ there are hairy solutions for which this timescale, for astrophysical black hole masses, is larger than the age of the Universe. The latter conclusion introduces the limited, but physically relevant concept of \textit{effective stability}. The former conclusion, allows us to identify an \textit{astrophysically viable domain} of such effectively stable hairy black holes, occurring, conservatively, for $M\mu \lesssim 0.25$. These are hairy BHs that form dynamically, from the superradiant instability of Kerr, within an astrophysical timescale, but whose own superradiant instability occurs only in a cosmological timescale.
\end{abstract}




\maketitle
\noindent{\bf {\em Introduction.}} 
The outstanding developments on observational black hole (BH) physics, including the detection of gravitational waves from binary BH mergers~\cite{Abbott:2016blz,Abbott:2016nmj,Abbott:2017vtc,Abbott:2017gyy,Abbott:2017oio} and the announced release of the first image of a BH by the Event Horizon Telescope collaboration~\cite{Lu:2014zja,Fish:2016jil}, make our epoch the most exciting time in their one hundred years history. The forthcoming  data opens up an unprecedented opportunity to test the true nature of BHs and, in particular, alternative models to the Kerr BH paradigm, together with their underlying physics~\cite{Loeb:2013lfa,Berti:2015itd,Cardoso:2016ryw,Johannsen:2015hib,Yunes:2016jcc}. 

On the theoretical front, viable BH models deserving phenomenological studies should pass at least three broad criteria: $(i)$ Appear in a well motivated and consistent physical model; $(ii)$ Have a dynamical formation mechanism; $(iii)$ Be sufficiently stable. In the last condition the key word is \textit{sufficiently}, as absolute stability is not mandatory for physical relevance; the latter always hinges on a competition of timescales. Amongst the many possible examples, an illustrative one in gravitational physics is the long term instability of the solar system~\cite{1994A&A...287L...9L}, due to 3-bodies interactions. Yet, the solar system \textit{exists} and planetary orbits may remain essentially unchanged during the lifetime of the Sun. This ineffectiveness of an instability that is known to be present in a physical system, within relevant timescales for that system, encodes a physically relevant sense of stability, that we call \textit{effective stability}.

Here we report on the emergence of effective stability in BH physics, within General Relativity. New families of stationary, asymptotically flat BHs in Einstein's theory, circumventing long standing ``no-hair" theorems~\cite{Bekenstein:1971hc,Bekenstein:1972ky,Bekenstein:1972ny,Herdeiro:2015waa} have been unveiled in the last few years, obeying condition $(i)$: Kerr BHs with synchronised hair~\cite{Herdeiro:2014goa,Herdeiro:2016tmi}. They have also been shown to obey condition $(ii)$~\cite{East:2017ovw,Herdeiro:2017phl}; in the vicinity of the Kerr limit,  these hairy BHs can form dynamically.  Very recently, however, these BHs have been argued to be unstable~\cite{Ganchev:2017uuo}. Here, we shall first clarify this instability is the anticipated superradiant instability~\cite{Herdeiro:2014jaa}. Then, we observe that the particular sample considered in~\cite{Ganchev:2017uuo} led to non-generic conclusions concerning timescales. In fact, results therein and other observations allow us to conservatively identify the domain of  hairy BHs for which this instability is not relevant for astrophysical BHs, within the age of the Universe. Thus, with respect to the superradiant instability, hairy BHs in this domain effectively obey condition $(iii)$ above.

\noindent{\bf {\em Kerr BHs with synchronised hair.}} The two basic examples of Kerr BHs with synchronised hair occur in Einstein's gravity minimally coupled to a massive, complex, scalar~\cite{Herdeiro:2014goa,Herdeiro:2015gia} or  vector~\cite{Herdeiro:2016tmi} field - see~\cite{Kleihaus:2015iea,Herdeiro:2015tia} for generalisations. They are asymptotically flat BH spacetimes, regular on and outside the horizon with a stationary and axisymmetric geometry. In spheroidal coordinates adapted to these symmetries, $(t,r,\theta,\phi)$, the metric has dependence $g^{(0)}_{\mu\nu}(r,\theta)$ whereas the matter field reads $\Phi^{(0)}=\varphi(r,\theta) \, e^{i(m\phi-\omega t)}$, where $\omega$ is the frequency,  $m\in \mathbb{Z}^+$ is the azimuthal harmonic index and  $\varphi(r,\theta)$ is a scalar (vector) profile function for the spin 0 (1) case. 

The defining physical property of these BHs is a synchronisation of the (non-zero) horizon angular velocity, $\Omega_H$, with the phase angular velocity of the matter field $\omega/m$: $\Omega_H=\omega/m$. The existence of the latter, which vanishes at the level of the matter energy momentum tensor, allow these solutions to circumvent well known no-hair theorems~\footnote{The first example of a BH with synchronised hair was found in five dimensional, asymptotically AdS space~\cite{Dias:2011at}.}, including classic ones by Bekenstein~\cite{Bekenstein:1971hc,Bekenstein:1972ky,Bekenstein:1972ny,Herdeiro:2015waa}.

\begin{widetext}

\begin{figure}[ht]
\begin{center}
\includegraphics[width=0.48\textwidth]{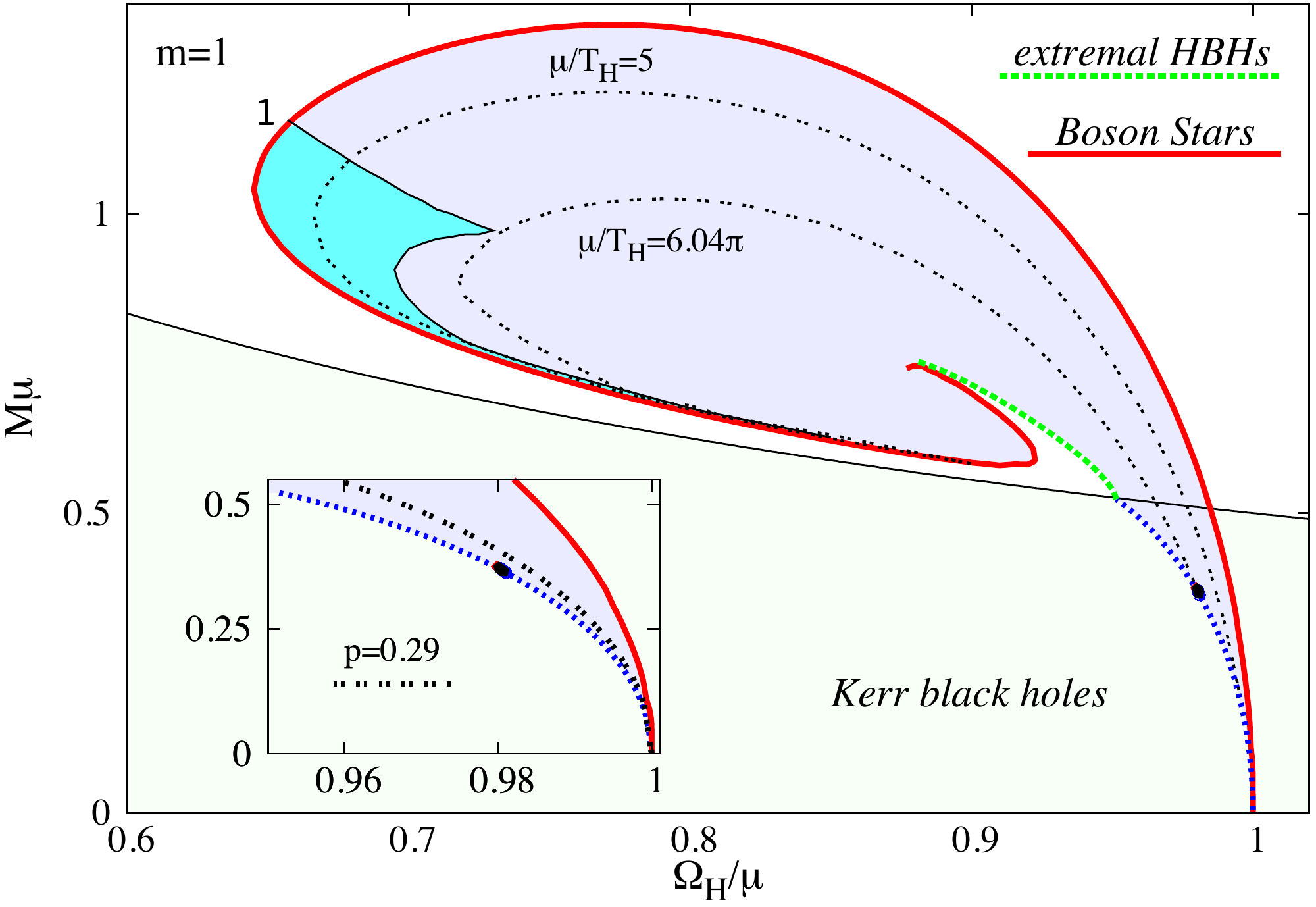}
\includegraphics[width=0.495\textwidth]{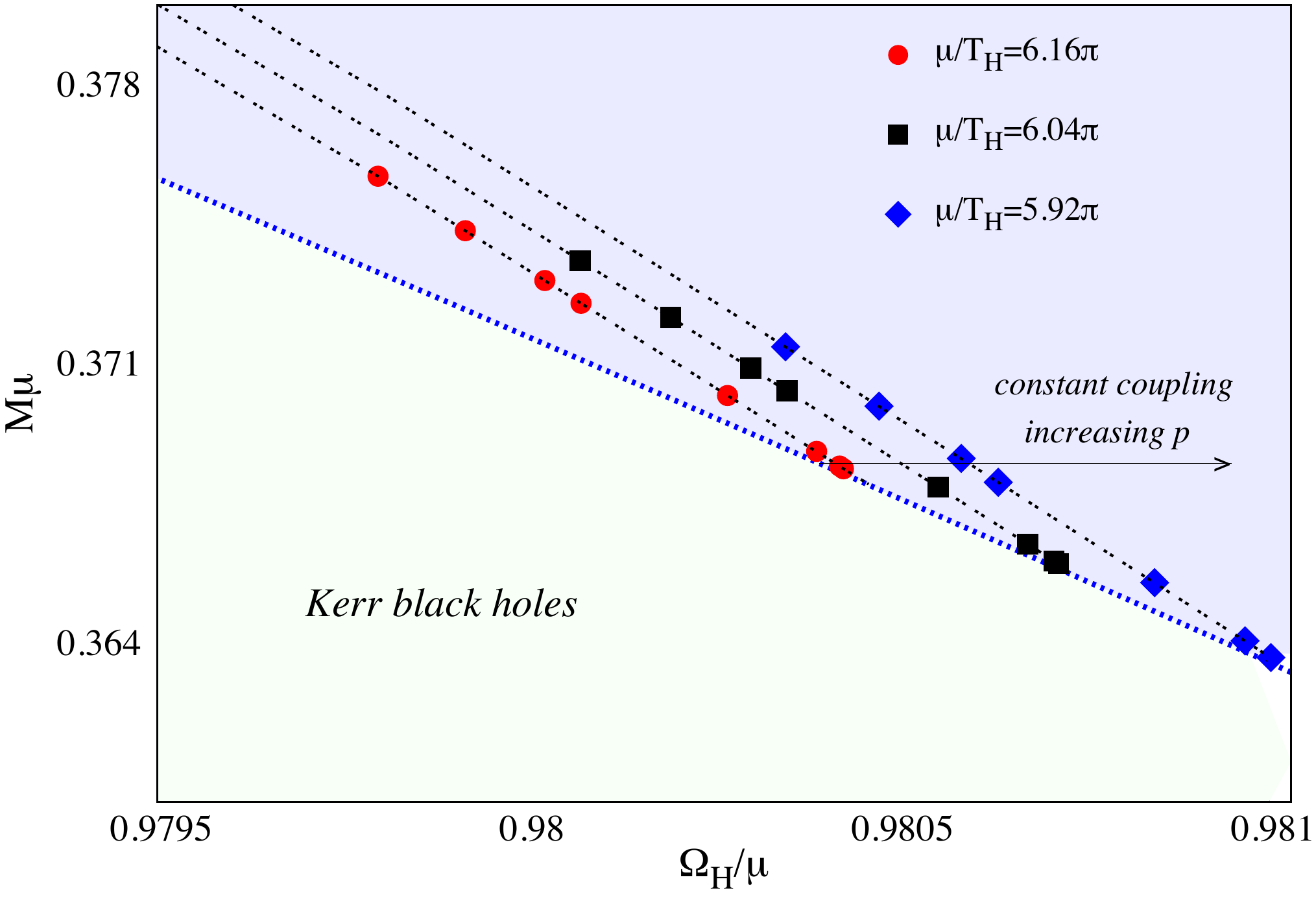}
\end{center}
\caption{\small (Left panel) Domain of existence of scalar hairy BHs with $m=1$. The domain (shaded blue region) is bounded by rotating boson stars (red solid line), Kerr BHs (blue dotted line -  the EL) and extremal hairy BHs (green dashed line). The wedge in bright blue is where ergo-Saturns exist; everywhere else hairy BHs have a Kerr-like ergo-region. Boson stars develop an ergo-torus at point ``1" as one moves towards the centre of the spiral. (Inset) Zoom in close to the EL.  (Right panel) Further zoom in of the black dot region, wherein all solutions in~\cite{Ganchev:2017uuo} lie: there are three sequences of solutions (red circles, black squares and blue diamonds), corresponding to the same colour code used in Fig. 1 therein. }
\label{fig1}
\end{figure}

\end{widetext}

The hairy BHs actually form a discrete set of families labeled by the integer $m\geqslant 1$~\footnote{There is another integer counting the number of nodes of the matter field and defining excited states. Herein we shall focus on the nodeless fundamental states.}. The domain of existence of the $m=1$ family in the scalar case is shown in Fig.~\ref{fig1} (main left panel). Hairy BHs interpolate between rotating boson stars~\cite{Schunck:2003kk} and a line of Kerr solutions that support zero modes ($i.e.$ with vanishing imaginary part) of the superradiant instability~\cite{Hod:2012px,Benone:2014ssa} - the~\textit{existence line} (EL). Analogous domains of existence can be constructed for both the vector case~\cite{Herdeiro:2016tmi,Herdeiro:2017phl} and  for higher values of $m$~\cite{Herdeiro:2014goa,Herdeiro:2015gia}.

\noindent{\bf {\em Existence of superradiant instabilities.}} 
BHs with synchronised hair, like Kerr, have an ergoregion~\cite{Herdeiro:2014jaa}. Fig.~\ref{fig1} shows the ergoregion structure in their domain of existence. In most of this domain the ergoregion is Kerr-like. But in the lower frequency part, there is a wedge where the ergoregion has more structure: it is an \textit{ergo-Saturn} composed of two disjoint parts, one of which being an ergo-torus inherited from the boson star environment. Boson stars only develop this ergo-torus at point ``1". In between this point and the maximal frequency boson stars are ergo-region free.

As discussed in~\cite{Herdeiro:2014jaa} and in~\cite{Herdeiro:2015gia} (Sec. 6),  superradiant instabilities triggered by perturbation modes with azimuthal harmonic index $\tilde{m}$ larger than that of the background should afflict the hairy BHs. Indeed, the existence of an ergoregion causes modes of a \textit{different} scalar (or other bosonic) field to be superradiantly unstable, in the appropriate frequency range.  This superradiant instability of hairy BHs was shown to occur for perturbations within the Einstein--Klein-Gordon model by Ganchev and Santos~\cite{Ganchev:2017uuo} (hereafter GS). Considering linear perturbation theory around the background $(g^{(0)},\Phi^{(0)}$), $g_{\mu\nu}=g^{(0)}_{\mu\nu}+h_{\mu\nu},$   $\Phi=\Phi^{(0)}+\eta$, GS argued that  a gauge exists wherein matter perturbations decouple from gravitational ones, and the matter perturbation equation becomes simply $
\Box^{(0)}\eta=\mu^2 \eta$~\footnote{An analogous decoupling of matter perturbations occurs in the case of Einstein-Yang-Mills theory and was used to study the stability of the corresponding hairy (known as \textit{coloured}) BHs~\cite{Volkov:1998cc}.}.

Facing this Klein-Gordon equation as that of a \textit{different} test scalar field on the background of  $m=1$ hairy BHs two conclusions follow: 1) the new scalar field, $\eta$, will have superradiant modes on the hairy BH background; and 2) for hairy BHs close to Kerr, such modes' timescales will be similar to those of the neighbouring Kerr BHs. This agrees with the GS findings, who reported the analysis of 24 scalar hairy BH solutions in the neighbourhood  corresponding to the black dot (left panel) in Fig.~\ref{fig1}~\footnote{The domain of existence display in Fig.~\ref{fig1} was obtained from a few thousands of numerical solutions. From our available data and that presented in~\cite{Ganchev:2017uuo} we identified the solutions used in the latter.}, in particular  obtaining for each case the dominant $\tilde{m}=2$ superradiant mode. We have checked that for all modes found therein, the real part of the frequency obeys $\mathcal{R}(\omega)\simeq 1.01 \Omega_H$, and thus
$ \Omega_H<\mathcal{R}(\omega)<2\Omega_H,$
which shows these $\tilde{m}=2$ modes are in the expected superradiant regime for $m=1$ hairy BHs. Moreover, for all these modes $
\mathcal{R}(\omega)<\mu$, ($cf.$ Fig.~2 in~\cite{Ganchev:2017uuo}) which shows these are quasi-bound states, with a radial function that decays exponentially at infinity.

\noindent{\bf {\em Timescales near the existence line.}} 
GS extracted two conclusions concerning the timescales of these superradiant instabilities. Firstly, that as the amplitude of the scalar cloud increases, the timescale of the instability grows with respect to that of a comparable Kerr BH (defined therein as having the same dimensionless spin $j\equiv J/M^2$ and \textit{temperature}); secondly, that the timescales involved, when applied to astrophysical BHs, are significantly smaller than the age of the Universe. A broader analysis will show both these conclusions can be circumvented, as we now discuss.

A central quantity in the study of superradiance is the coupling $M \mu$. It is well known that the strength of the instability - measured by the imaginary part of the frequency, $\mathcal{I}(\omega)$ -  is highly sensitive to the value of the coupling~\cite{Brito:2015oca}. For Kerr, the superradiant instability triggered by a scalar field is strongest for the $\tilde{m}=\ell=1$ mode  (with angular quantum number $\ell$) when $M\mu\sim 0.42$, for $j\sim 0.99$, whereas for $j\sim 0.95$ the instability is maximal at $M\mu\sim 0.343$~\cite{Dolan:2007mj}. Away from this maximal efficient value, the strength of the instability swiftly decreases as $\sim (M\mu)^{4+4\ell}$ for $M\mu\ll 1$~\cite{Detweiler:1980uk} and $10^{-7}e^{-3.7 M\mu}$ as $M\mu\gg 1$~\cite{Zouros:1979iw,Arvanitaki:2010sy}.

All data in~\cite{Ganchev:2017uuo} have $M \mu\sim 0.36-0.38$, $j\simeq 0.95-0.97$ and are thus close to the maximal efficiency coupling in Kerr. Moreover these data refer to rather Kerr-like hairy BHs: for all solutions reported, the parameter $p$~\cite{Herdeiro:2017phl}, measuring the fraction of total energy stored as scalar field energy outside the horizon, is $p\lesssim 0.02$. In this \textit{feeble hair} regime, the expectation that the timescales of the superradiant instability are of the same order of magnitude as those of a massive scalar field on a comparable Kerr BH is confirmed. This justifies estimating the timescales of the superradiant instabilities of hairy BHs in other equally feeble hair regions using Kerr timescales. From the discussion in the previous paragraph, it is clear the timescale of the instability will decrease rapidly with the coupling.

\begin{figure}[ht]
\begin{center}
\includegraphics[width=0.48\textwidth]{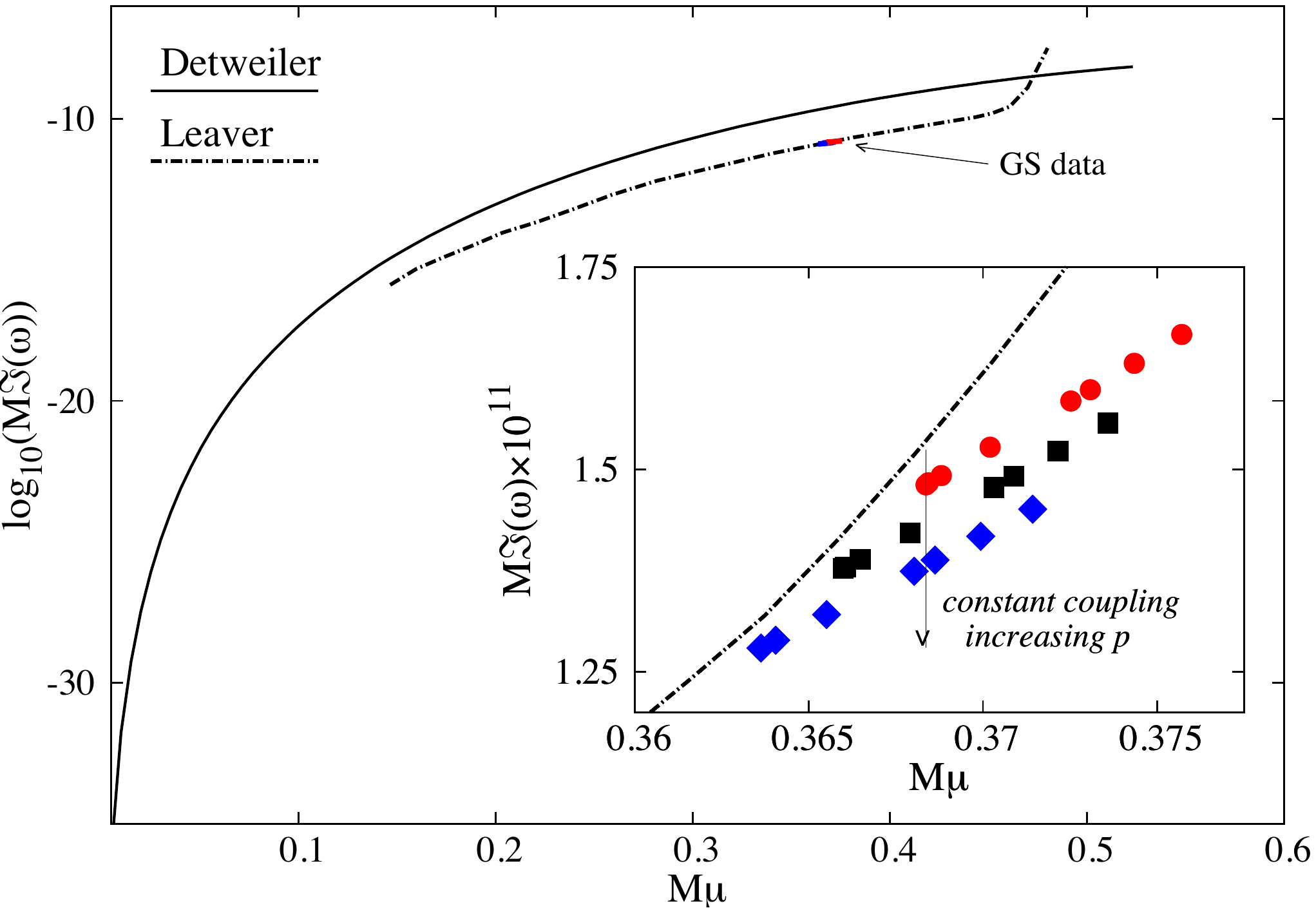}
\end{center}
\caption{\small Strength of the instability of the $\tilde{m}=2$ mode along the $m=1$ EL. The dashed (solid) curve was obtained  through Leaver's method (Detweiler's approximation). The GS data is shown in the main panel (falling on top of the dashed curve) and in the inset.}
\label{fig2}
\end{figure}

In Fig.~\ref{fig2} we present the strength of the instability for the dominant $\tilde{m}=2$ superradiant modes of a massive scalar field on the Kerr BHs that lie along the $m=1$ EL (blue dotted line in Fig.~\ref{fig1}). %
These results (dashed curve), obtained using Leaver's method~\cite{Leaver:1985ax}, are in agreement with previous literature ($e.g.$~\cite{Dolan:2007mj,Dolan:2012yt}).
 Fig.~\ref{fig2} also shows Detweiler's approximation~\cite{Detweiler:1980uk}  (solid curve), valid for $M\mu\ll 1$, which overestimates the strength of the instability by about one order of magnitude around $M\mu\simeq 0.14$, the lowest coupling at which numerical results from Leaver's method can be obtained. The figure also shows the instability strengths for the hairy solutions in GS, which, in the inset, can be seen to fall \textit{below} this curve, in fact emerging from it as the sequence of hairy BHs approaches the Kerr limit ($cf.$ Fig.~\ref{fig1}, right panel).

As anticipated above, decreasing $M \mu$ the strength of the instability is rapidly suppressed. While the sample in~\cite{Ganchev:2017uuo} has $M\mathcal{I}(w)\sim 10^{-11}$, for $M \mu\lesssim 0.25$ we obtain  $M\mathcal{I}(w)\lesssim 1.13\times10^{-13}$, an instability two orders of magnitude weaker.  Such a decrease in the strength is enough to provide an example wherein the instability is ineffective within the age of the Universe, $\tau_{\rm U}$: for a supermassive BH with $10^{10} \, M_\odot$ the corresponding timescale is precisely of the order of $\tau_{\rm U}\eqsim 4.35\times10^{17}\, s$.

\noindent{\bf {\em Beyond the feeble hair regime.}} 
Leaving the proximity of the EL, the relevant question is if, as the BHs become hairier, the strength of the superradiant instability increases. In GS, this analysis was done along three sequences of constant temperature. For this choice and for increasing $j$, the strength of the Kerr instability decreases. For fixed coupling, on the other hand, the strength of the instability increases for faster spinning Kerr BHs~\footnote{This trend is only violated very close to extremality, around $j\sim 0.997$~\cite{Dolan:2012yt}.}. The underlying reason for the former behaviour is that keeping constant temperature $T_H$ (in units of $\mu$), implies, for Kerr BHs, that the coupling $\mu M$ decreases with increasing $j$, since $T_H=\{4\pi M[1+(1-j^2)^{-1/2}]\}^{-1}$. This decrease of the coupling turns out to dominate the trend. For the hairy BHs sequence at constant $T_H$, on the other hand, the coupling increases with increasing $j$, corresponding to moving away from the EL  - Fig~\ref{fig1} (black double dotted lines). Thus, fixing the temperature one is increasing the ratio of the coupling between hairy and Kerr BHs, as $j$ increases, which is why the relative strength of the instability increases.

Rather than analysing sequences of hairy BHs at constant temperature, which appear to spiral along the domain of existence ($cf.$ Fig.~\ref{fig1}, left panel), we consider a sequence at fixed coupling $M \mu$; these are simply horizontal lines in Fig.~\ref{fig1}. Then, initiating the sequence at the EL, one starts with a Kerr BH and gradually increases the amount of hair, making the parameter $p$ vary from $0$ to $1$. The sequence ends in a boson star \textit{without} ergoregion, thus without superradiant instabilities. The trend in the GS data along this sequence is a \textit{decrease} of the instability strength, in contrast with the lines of constant $T_H$. This can be easily established by comparing the right panel of Fig.~\ref{fig1} and the inset of Fig.~\ref{fig2}. Moving away from the EL, at constant coupling, corresponds to a rightwards horizontal line in the former and downwards vertical line in the latter.

The trend at fixed coupling was anticipated in~\cite{Herdeiro:2014jaa} using a naive measure dubbed ``ergosize", $\Delta a$. This is defined as the area of the ergo-sphere minus the area of the event horizon, with a normalisation factor.
$\Delta a$ was observed to correlate (directly) with the strength of the instability in various examples at fixed $\mu M$. For the hairy BHs it decreases along constant coupling sequences starting at the EL, hinting at a suppression of the instability. Another suggestive fact is that, along the same sequences, $j_H$, defined as $j$ but in terms of \textit{horizon} quantities~\cite{Herdeiro:2015moa}, decreases: the horizon of hairier BHs carries less dimensionless spin~\footnote{Along the constant coupling sequences, starting from the EL, $j_H$ decreases but $j$ increases. Thus, an increasing proportion of the dimensionless spin is stored in the hair. Along the $T_H$=constant sequences in~\cite{Ganchev:2017uuo}, on the other hand, both $j_H$ and $j$ increase.}. These observations are consistent with the total quenching of the instability, along these sequences, in the boson star limit  (no ergo-region), and suggest that solutions close to the boson star line are also effectively stable against superradiance. In this region, however, there is no known dynamical formation mechanism, which is only known in the vicinity~\footnote{Dynamical formation from Kerr, via superradiance, yields hairy BHs that obey the thermodynamical limit of $p\lesssim 0.29$. The corresponding region lies between the EL and the double dotted line in the inset of Fig.~\ref{fig1} (left).} of the EL~\cite{East:2017ovw} (see also~\cite{Herdeiro:2017phl}).

\noindent{\bf {\em Astrophysically viable domain: stability bound.}} 
The discussion above supports the conclusion that, for fixed $M\mu$, the leading $\tilde{m}=2$ superradiant mode of the \textit{Kerr solutions} along the EL of Fig.~\ref{fig1},  gives a \textit{lower bound} on the instability timescale, $\tau_{\rm EL}^{\tilde{m}=2}(M\mu)$ (upper bound on the instability strength) of the leading ($\tilde{m}=2$) superradiant mode of the $m=1$ hairy BHs. 

\begin{figure}[ht]
\begin{center}
\includegraphics[width=0.48\textwidth]{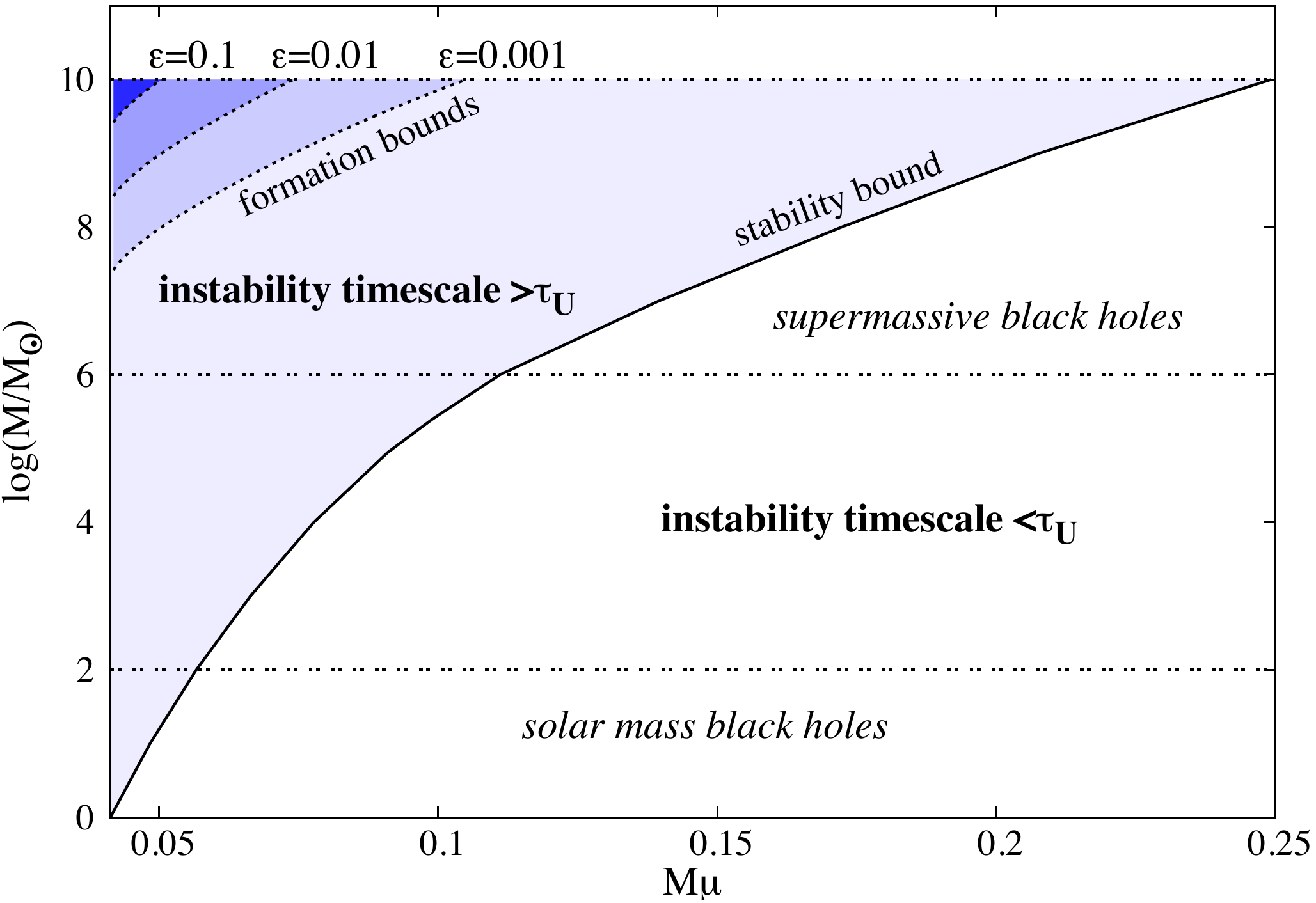}
\end{center}
\caption{\small Astrophysically viable domain of hairy BHs (shaded regions). For BHs with Log$(M/M_\odot) =  (0,2,6,10)$, the stability bound (solid curve) yields $M\mu\sim (0.04,0.06,0.11,0.25)$. The dotted curves are formation bounds within $\tau=\epsilon \tau_{\rm U}$.}
\label{fig3}
\end{figure}

Using, thus,  $\tau_{\rm EL}^{\tilde{m}=2}(M\mu)$ as a conservative bound on the superradiant instability of hairy BHs with a given $M \mu$ we can delimit the astrophysically viable domain of these BHs, identifying the range of masses, for a given coupling, wherein these hairy BHs are effectively stable.  Requiring $\tau_{\rm EL}^{\tilde{m}=2}(M\mu) \sim \tau_{\rm U}$ and using the data in Fig.~\ref{fig2}, matching the Detweiler's approximation for $M\lesssim 0.1$ to the result from Leaver's method,  we translate the requirement $\tau_{\rm EL}^{\tilde{m}=2}(M\mu) \sim \tau_{\rm U}$ into a relation $M/M_\odot(M\mu)$ - the solid curve in Fig.~\ref{fig3}. Points \textit{above} this curve describe BH masses and couplings for which the instability timescale is larger than $\tau_{\rm U}$ and hence \textit{effectively stable} against superradiance. We emphasise this analysis is very conservative: firstly because we are taking the strength of the instability at the largest possible value for that coupling, rather than the actual (smaller) value for the hairy BH; secondly, because we are putting the threshold for the effective stability at $\tau_{\rm U}$; but even one order of magnitude below yields a sufficiently long timescale so that the BHs may be relevant astrophysically. Still, the unavoidable conclusion is that there is a domain of effectively stable hairy BH solutions (against superradiance).

\noindent{\bf {\em Astrophysically viable domain: formation bound.}} 
It remains to discuss whether the domain of effectively stable hairy BHs includes those that may form dynamically from the superradiant instability of Kerr.  The latter form hairy BHs due to the leading superradiant instability, triggered by the fundamental $\tilde{m}=1=\ell$ superradiant mode of a massive scalar field on Kerr. According to~\cite{East:2017ovw} this process is approximately adiabatic and thus the evolution occurs along a constant coupling line and for constant total angular momentum. Both mass and angular momentum are transferred to the hair with no significant losses; thus the global quantities are the same for the Kerr progenitor and the hairy descendent. To evaluate the effectiveness of the process, we recall that the fastest instability occurs for nearly extremal Kerr BHs, with $j\sim 0.99$, regardless of $M\mu$~\cite{Dolan:2007mj}. Requiring the corresponding timescale to be a fraction of the age of the Universe, $\tau^{\tilde{m}=1}_{j=0.99}(M\mu)=\epsilon \tau_{\rm U}$, and using the data of~\cite{Dolan:2007mj}, we again obtain a curve in the diagram of Fig.~\ref{fig3}. For a given coupling, only points \textit{below} the curve describe BHs that form within the considered fraction of $\tau_{\rm U}$. In Fig.~\ref{fig3}, three such (dotted) curves are illustrated, for $\epsilon=0.1,0.01,0.001$. As expected, requiring a smaller fraction of $\tau_{U}$ as the formation bound, excludes a larger region of the viable domain. Thus, the light blue shaded regions in Fig.~\ref{fig3} correspond to couplings and physical masses of hairy BHs that form, from the superradiant instability of Kerr, faster than one thousandth of the age of the Universe, but whose own dominant superradiant instability has a timescale larger than $\tau_{U}$.

\noindent{\bf {\em Conclusion.}} 
The advance~\cite{Ganchev:2017uuo} in the understanding of the superradiant instabilities of the hairy BHs found in~\cite{Herdeiro:2014goa} paved the way to the present observations that these instabilities are quenched in some regions of the parameter space for astrophysical BH masses. Hairy BHs are afflicted by the same instability that exists for Kerr BHs in the presence of ultralight bosonic fields, and that can lead to their dynamical formation~\cite{East:2017ovw,Herdeiro:2017phl}. The instability of the hairy BH descendent, however, is orders of magnitude weaker than that of the Kerr progenitor, since higher $\tilde{m}$ modes have longer timescales.  We have identified, conservatively, an astrophysically viable domain of BHs with synchronised scalar hair. A key question, already under scrutiny, is whether the astrophysical phenomenology in the dynamically allowed region is also viable.

Finally, what is the fate of the hairy BHs that are \textit{not} effectively stable? The instability will grow more hair, of the $\tilde{m}$ mode, and interference with the background $m$ mode will induce dissipation (gravitational radiation emission). The spacetime, likely, migrates to a higher $m$ (lower $M,J$) hairy BH, repeating the process until effective stability is achieved. 

\bigskip

\noindent{\bf {\em Note Added.}} 
After this paper appeared in the arXiv, a revised version of~\cite{Ganchev:2017uuo} also appeared. The revised version acknowledged that hairy black holes are less unstable against superradiance than comparable Kerr black holes, in agreement with the results described in this paper.

\bigskip  

\noindent{\bf {\em Acknowledgements.}}
We would like to thank V. Cardoso, P. Cunha and N. Sanchis-Gual for comments on a draft of this paper and S. Dolan for sharing some of the data in~\cite{Dolan:2007mj} with us. J. C. D.  acknowledges support from DGAPA-UNAM through grant IA101318. C. H. and E. R. acknowledge funding from the FCT-IF programme.  This project has received funding
from  the  European  Union's  Horizon  2020  research  and  innovation  programme  under  the H2020-MSCA-RISE-2015 Grant No.   StronGrHEP-690904, the H2020-MSCA-RISE-2017 Grant No. FunFiCO-777740  and  by  the  CIDMA  project UID/MAT/04106/2013. 
The authors  would also  like  to  acknowledge networking support by the COST Action GWverse CA16104. 

\bigskip


\bibliography{letter_es}

 
\end{document}